\documentclass[pra,preprint,superscriptaddress,showpacs,preprintnumbers,amsmath,amssymb,eqsecnum]{revtex4}

\usepackage{amsmath,amssymb}
\usepackage[dvipsnames]{xcolor}

\usepackage{epsfig}
\usepackage{dcolumn}
\usepackage{bm}

\begin{document}

\title{Wigner function and the successive measurement of position and momentum 
}

\author{Pier A. Mello}
\address{Departamento de Sistemas Complejos, Instituto de
F\'{\i}sica, Universidad Nacional Aut\'{o}noma de M\'{e}xico,
M\'{e}xico, D.F. C.P. 04510}

\author{Michael Revzen}
\address{Department of Physics, Technion - Israel Institute of Technology, Haifa 32000, Israel
}

\date{\today}

\begin{abstract}

Wigner function is a ``quasi-distribution"
that provides a representation of the state of a quantum mechanical system in the phase space of position and momentum.
In this paper we find a relation between Wigner function and appropriate measurements involving the system position and momentum which generalize the von Neumann model of measurement.
We introduce two {\em probes} coupled successively in time to projectors associated with the system position and momentum.
We show that one can relate Wigner function to Kirkwood joint quasi-distribution of position and momentum, the latter, in turn, being a particular case of successive measurements.
We first consider the case of a quantum mechanical system described in a continuous Hilbert space, and then turn to the case of a discrete, finite-dimensional Hilbert space.

\end{abstract}


\maketitle

\section{Introduction}
\label{intro}

Wigner function was originally introduced to provide a phase-space representation of the state of a Quantum-Mechanical system described in a continuous Hilbert space \cite{wigner}.
Wigner function is termed a ``quasi-distribution", as it may become negative in some portions of phase space \cite{leonhardt,schleich}.
Indeed, as is well known, Quantum Mechanics (QM) precludes a proper joint probability distribution of position $q$ and momentum $p$. 
However, in many respects Wigner function plays a role similar to the phase space distribution function in classical statistical mechanics \cite{leonhardt,schleich,khanna-mello-revzen};
therefore, we find it natural to inquire whether one can relate it to appropriate measurements involving the position and momentum of the system.

The idea we shall develop is to introduce the first stage of the measurement, or ``pre-measurement", explicitly in the QM description, by coupling, successively in time,
the system observables we wish to study to auxiliary degrees of freedom, or {\em ``probes"}, and detect the probes, not the system itself.
This procedure represents a generalization of von Neumann's model of measurement \cite{vonNeumann,lars07,lars-mello}.
Specifically, we shall couple projectors associated with the system position and momentum to two independent probes, at times $t_1$ and $t_2$, respectively.
It turns out that one can relate Wigner function to correlations of observables 
--each belonging to one of the two probes-- 
which, being distinct degrees of freedom and external to the system,
{\em are compatible} and admit a joint probability distribution.
These probe correlations are thus experimentally accessible.

We shall first relate Wigner's function to the so-called Kirkwood joint quasi-probability distribution of position and momentum \cite{kirkwood,dirac} which is, in general, a complex quantity.
It is then Kirkwood's distribution which can be expressed in terms
of the above mentioned probe correlations, in the limit in which the coupling becomes very weak.

In a historical context, it is interesting to mention that Kirkwood introduced the joint quasi-probability distribution in phase space that bears his name a year later than Wigner introduced his, and with similar motivations related to statistical mechanics applications.
In the next decade Dirac introduced essentially the same joint quasi-probability distribution for non-commuting observables, with the aim of ``discussing trajectories for the motion of a particle in QM".

We should remark that in the field of Quantum Optics, Wigner function has been related to a set of measurable quantities different from the ones considered in the present paper, namely, quadrature distributions which are experimentally available,
a method that constitutes an application to QM of the computer-aided tomography scan \cite{leonhardt,schleich,khanna-mello-revzen}.

Other quasi-distributions have been proposed in the literature: 
e.g., Ref. \cite{cohen} presents a family of 
quasi-distribution functions, of which Wigner function
--which is the distribution considered herewith--
is a special case.
As for the relation between Wigner's function and Kirkwood's
quasi-distribution, we also refer the reader to 
Refs. \cite{berge89,lee95,moya-et-al08}.

The paper is organized as follows.
In the next section we develop the scheme we just outlined, for a Quantum-Mechanical system described in a continuous Hilbert space.
In Sec. \ref{wigner-kirkwood discrete} we then turn to studying a discrete, finite-dimensional Hilbert space.
The notion of Wigner function for a discrete Hilbert space is a topic which has been widely studied in the literature (a selection of these contributions is represented by Refs. \cite{buot74,hannay-berry80,cohen-scully86,wootters87,galetti-de-toledo-piza88,cohendet-et-al88,kasperkovitz-peev94,leonhardt95-96,luis-perina98,rivas-ozorio99,bandyopa,gibbons-et-al04,vourdas04,micha-WF012}).
Here, we adopt an alternative definition
--which will be of interest in a geometrical context to be described elsewhere--
as the starting point to develop the scheme presented above.
We shall see that the discrete case is free from a number of divergences that are encountered in the continuous case.
Various specific algebraic calculations have been carried out in appendices, in order not to interrupt the main presentation.
We finally conclude in Sec. \ref{concl}.

\section{Wigner function and Kirkwood quasi distribution for a continuous Hilbert space}
\label{wigner-kirkwood continuous}

\subsection{The Wigner transform of an operator defined in a continuous Hilbert space}

The Wigner transform (WT) of an operator $\hat{A}$ is a mapping from Hilbert space to phase space \cite{wigner}. 
It can be expressed as the inverse Fourier transform of the characteristic function of the operator. 
Using units in which $q$ and $p$ are dimensionless, and $\hbar=1$, we have the definition \cite{leonhardt,schleich,khanna-mello-revzen}
\begin{subequations}
\begin{eqnarray}
W_{\hat{A}}(q,p)
&=& \frac{1}{2\pi} \int_{-\infty}^{\infty} \int_{-\infty}^{\infty}
\tilde{W}_{\hat{A}}(u,v)e^{i(uq + vp)} du dv  
\label{WF continuous a}
\\
\tilde{W}_{\hat{A}}(u,v) 
&=& {\rm Tr} \left[\hat{A} e^{-i(u\hat{q} + v\hat{p})} \right]  .
\label{WF continuous b}
\end{eqnarray}
\label{WF continuous}
\end{subequations}
When the operator $\hat{A}$ is the density operator $\hat{\rho}$, we speak of its WT as the Wigner function (WF) of the state.
The definition (\ref{WF continuous}) is equivalent to the standard one, presented, for convenience, in Eq. (\ref{WT,continuous,standard}).

The WT of an operator $\hat{A}$ can also be expressed as
\begin{equation}
W_{\hat{A}}(q,p) = {\rm Tr} [\hat{A} \hat{P}(q,p)],
\label{WT=tr(AP)}
\end{equation}
$\hat{P}(q,p)$ being a Hermitean operator.
Using the definition of WT given in Eq. (\ref{WT,continuous,standard}), $\hat{P}(q,p)$ can be written as
\begin{equation}
\hat{P}(q,p)
= \int_{-\infty}^{\infty}e^{-ipy} 
\left|q-\frac{y}{2}\right\rangle \left\langle q+\frac{y}{2}\right| dy \; .
\label{P(q,p) continuous 1}
\end{equation}
We can also use the {\em mutually unbiased bases} \cite{khanna-mello-revzen} 
(MUB) states 
$|x', \theta \rangle$, eigenstates of the operator 
$\hat{X}_{\theta} = \hat{q}\cos\theta + \hat{p}\sin \theta$ 
--and hence eigenstates of the exponential operator appearing in 
Eq. (\ref{WF continuous b}))--
which satisfy the eigenvalue equation 
$\hat{X}_{\theta}|x', \theta \rangle = x' |x', \theta \rangle$,
to express the operator $\hat{P}(q,p)$ as
\begin{subequations}
\begin{eqnarray}
\hat{P}(q,p)
&=& \frac{1}{2 \pi}
\int_{0}^{\pi} d\theta \int_{-\infty}^{\infty} dx' \int_{0}^{\infty} dt
|t| e^{-it(x'- q\cos\theta - p\sin \theta)}
|x';\theta\rangle \langle x';\theta| ,
\label{P(q,p) continuous mub a} \\
&=&-\frac{1}{\pi}{\cal P}
\int_{0}^{\pi} d \theta
\int_{-\infty}^{\infty} dx'
\frac{\frac{\partial}{\partial x'} |  x'; \theta \rangle \langle x'; \theta|}{x'-(q\cos\theta + p \sin\theta)} \; .
\label{P(q,p) continuous mub b}
\end{eqnarray}
\label{P(q,p) continuous mub}
\end{subequations}
(Cf. Ref. \cite{khanna-mello-revzen}, Eq. (A6)
(where $\rho(x,y)$ is to be identified with $W_{\hat{\rho}}(q,p)$ and
$\rho_{\theta}(x')$ with $\langle x', \theta|\hat{\rho}|x', \theta\rangle$),
and Eq. (23)).

The operator $\hat{P}(q,p)$ and the WT of an (arbitrary) operator $\hat{A}$ possess the following attributes:

1) The matrix elements of the operator $\hat{P}(q,p)$ of 
Eq. (\ref{P(q,p) continuous 1}) in the coordinate basis are given by
\begin{equation}
\langle q|\hat{P}(q',p')|\bar{q}\rangle
= e^{ip'(q-\bar{q})}
\delta(q+\bar{q}-2q') \; .
\label{m. els. P{q,p} cont}
\end{equation}

2) The WT of a Hermitean operator $\hat{A}$ is real, which follows immediately from the Hermiticity of $\hat{P}(q,p)$.

3) The operators $\hat{P}(q,p)$ fulfill the following orthogonality and closure relations
\begin{subequations}
\begin{eqnarray}
\frac{1}{2\pi} {\rm Tr}\left[\hat{P}(q,p) \hat{P}(q',p')\right]
&=& \delta (q-q') \delta (p-p') \; ,  
\label{orthog Ps cont} \\
\frac{1}{2\pi}  
\int_{-\infty}^{\infty}  \int_{-\infty}^{\infty}\hat{P}(q,p) dq dp 
&=& \mathbb{I} \; ,
\label{closure cont.}
\end{eqnarray}
\label{orthog. and closure of Ps continuous}
\end{subequations}
$\mathbb{I}$ being the unit operator.

4) The WT of the operators $\hat{A}$ and $\hat{B}$ satisfy the ``product formula", or ``overlap formula" (see, e.g., Ref. \cite{leonhardt}, Eq. (3.5), and  Ref. \cite{schleich}, Eq. (3.20))
\begin{equation}
\int_{-\infty}^{\infty}  \int_{-\infty}^{\infty} 
W_{\hat A}(q,p)   W_{\hat B}(q,p) \frac{dq dp}{2 \pi}
= {\rm Tr}(\hat A \hat B) \; .
\label{product formula continuous}
\end{equation}

5) The WF for the state $\rho$ satisfies the marginality relation
\begin{eqnarray}
{\rm Tr}(\hat{\rho} \hat{\mathbb{P}}_{x'}^{\theta})
&=& \langle x', \theta |\hat{\rho} |x', \theta \rangle 
\nonumber \\
&=& \int_{-\infty}^{\infty}  \int_{-\infty}^{\infty} 
W_{\hat \rho}(q,p)   \delta(x' - (q \cos\theta + p \sin \theta)) 
\frac{dq dp}{2 \pi} \; ,
\label{marginality cont}
\end{eqnarray}
(see Ref. \cite{khanna-mello-revzen}, Eq. (22))
which states that if the system is in state $\hat{\rho}$, the probability to find it in the pure state $|x', \theta \rangle$  is given by the integral of the WF along the line
$q \cos\theta + p \sin \theta = x'$ in phase space. 
In particular, the marginal probability of $q$ and that for $p$ take the standard form.
Expression (\ref{marginality cont}) is referred to as the {\em Radon transform} \cite{leonhardt,schleich,khanna-mello-revzen} of the Wigner function $W_{\hat \rho}(q,p)$.

6) The WF is normalized as
\begin{equation}
\int_{-\infty}^{\infty}  \int_{-\infty}^{\infty} 
W_{\hat \rho}(q,p) \frac{dq dp}{2 \pi}
= 1 \; .
\label{normalization}
\end{equation}

\subsection{Relation between Wigner function and Kirkwood quasi-distribution for a continuous Hilbert space}

As shown in \ref{derivation of W(K) continuous}, one can express Wigner function in terms of Kirkwood's quasi-distribution as
\begin{equation}
W_{\hat{\rho}}(q,p)
= 2 \int \int dq' dp' e^{2i(q-q')(p-p')} K(p',q').
\label{W(K) continuous 1}
\end{equation}
Here, the quantity
\begin{subequations}
\begin{eqnarray}
K(p,q) &=& {\rm Tr}(\hat{\rho} \; \hat{\mathbb{P}}_{p} \hat{\mathbb{P}}_{q}),
\label{K continuous 1}
\end{eqnarray}
with the definition
\begin{eqnarray}
\hat{\mathbb{P}}_{q} &=& |q\rangle \langle q| 
\label{Pq, continuous} \\
\hat{\mathbb{P}}_{p} &=& |q\rangle \langle p|,
\label{Pp, continuous}
\end{eqnarray}
\label{Kpq, Pq, Pp continuous}
\end{subequations}
is Kirkwood's joint quasi-distribution \cite{kirkwood,dirac} of $q$ and $p$, which is, in general, complex.
Similar results can be found in Refs. \cite{berge89,lee95,moya-et-al08}.

The operators $\hat{\mathbb{P}}_q$ and $\hat{\mathbb{P}}_p$ are not proper position and momentum projectors, since they are not idempotent.
In order to use the formalism developed in Ref. \cite{lars-mello} we use, instead, the operators $\hat{\mathbb{P}}_{q_n}$ and $\hat{\mathbb{P}}_{p_m}$ defined in 
App. \ref{vnM x,p}.
For this purpose, we write Eq. (\ref{W(K) continuous 1}) as
\begin{subequations}
\begin{eqnarray}
W_{\hat{\rho}}(q,p)
&=& 2 \sum_{n,m = - \infty}^{\infty} 
\int_{q_n-\delta/2}^{q_n+\delta/2} dq' 
\int_{p_m-\delta/2}^{p_m+\delta/2} dp' 
e^{2i(q-q')(p-p')} 
{\rm Tr}_s \left(\hat{\rho}_s \; \hat{\mathbb{P}}_{p'} \hat{\mathbb{P}}_{q'}\right)
\label{W(K) 2 a}   \nonumber \\ \\
&\approx& 2 \sum_{n,m = - \infty}^{\infty}
e^{2i(q-q_n)(p-p_m)}
{\rm Tr}_s\left(\hat{\rho}_s  
\int_{p_m-\delta/2}^{p_m+\delta/2} dp' \; \hat{\mathbb{P}}_{p'} 
\int_{q_n-\delta/2}^{q_n+\delta/2} dq' \; \hat{\mathbb{P}}_{q'}\right)
\label{W(K) 2 b}  \nonumber \\ \\
&=& 2 \sum_{n,m = - \infty}^{\infty}
e^{2i(q-q_n)(p-p_m)} 
{\rm Tr}_s \left(\hat{\rho}_s  \hat{\mathbb{P}}_{p_m}\hat{\mathbb{P}}_{q_n} \right)
\label{W(K) 2 c}   \\
&=& 2 \sum_{n,m = - \infty}^{\infty}
e^{2i(q-q_n)(p-p_m)} K(p_m,q_n) \; ,
\label{W(K) 2 d}
\end{eqnarray}
\label{W(K) 2}
\end{subequations}
where $K(p_m,q_n)$ is Kirkwood's joint quasi-probability distribution of $p_m$ and $q_n$ defined in Eq. (\ref{K(pm,qn)}).
The discretization involved in going from Eq. (\ref{W(K) 2 a}) to (\ref{W(K) 2 b}) is an approximation. We expect that approximation to be justified if the interval $\delta$ is small enough that the factor $e^{2i(q-q')(p-p')}$ does not vary appreciably for $q'$ and $p'$ inside that interval.
Alternatively, it could be justified using a ``mean-value theorem" \cite{apostol}.
An argument where the approximation appears at the level of $c$-number functions can be found in App. \ref{vnM x,p}, right below Eq. (\ref{Ppm}).

According to Eq. (\ref{K(pm,qn) 1}), Kirkwood's distribution, in turn, can be expressed in terms of the position-position correlation of the two probes and their momentum-position correlation:
these are compatible variables, detected in measurements described by von-Neumman's model with very weak coupling;
specifically, in this model 
the observables coupled in succession to the two probes are the 
projectors for position and momentum of the system proper.
Substituting the result of Eq. (\ref{K(pm,qn) 1}) in Eq. (\ref{W(K) 2}) we thus find
\begin{eqnarray}
W_{\hat{\rho}}(q,p)
&=& 2 \sum_{n,m = - \infty}^{\infty}
e^{2i(q-q_n)(p-p_m)} 
\nonumber \\
&&\times \left\{
{\rm lim}_{\epsilon_1 \to 0} \frac{1}{\epsilon_1 \epsilon_2}
\left[\langle \hat{Q}_1 \hat{Q}_2 \rangle^
{(\hat{\mathbb{P}}_{p_m} \leftarrow \hat{\mathbb{P}}_{q_n})} 
+ \frac{i}{2 \sigma_{P_1}^2} 
\langle \hat{P}_1 \hat{Q}_2 \rangle
^{(\hat{\mathbb{P}}_{p_m} \leftarrow \hat{\mathbb{P}}_{q_n})} 
\right]
\right\}.
\nonumber \\
\label{W(probe correl)continuous}
\end{eqnarray}
This result states that Wigner function, which is defined in the system phase space, can be related to a set of {\em measurable quantities}, consisting of the two-probe correlations detected in the experimental setup described above, and thereby reconstructed therefrom.

\section{Wigner function and Kirkwood quasi distribution for a discrete, finite-dimensional Hilbert space}
\label{wigner-kirkwood discrete}

The analysis performed in the previous section for a continuous Hilbert space
will now be extended with a similar philosophy to a discrete, finite-dimensional Hilbert space.

\subsection{The Wigner transform for a discrete, finite-dimensional Hilbert space}
\label{}

The possibility of defining a WT for a Hilbert space of finite dimensionality
has been studied extensively in the literature  \cite{buot74,hannay-berry80,cohen-scully86,wootters87,galetti-de-toledo-piza88,cohendet-et-al88,kasperkovitz-peev94,leonhardt95-96,luis-perina98,rivas-ozorio99,bandyopa,gibbons-et-al04,vourdas04,micha-WF012}.
Here we propose, for the WT of an operator $\hat{A}$
defined in a Hilbert space of dimensionality $N$, the definition
\begin{subequations}
\begin{eqnarray}
W_{\hat{A}}(q,p)
&=&\frac{1}{N}
\left\{
\sum_{b=0}^{N-1}\sum_{k=1}^{N-1}
\tilde{W}_{\hat{A}}(k,b)
e^{i\frac{2\pi}{N}k(-p + b q)}
+ \sum_{l=0}^{N-1}
\tilde{W}_{\hat{A}}(l) e^{i\frac{2\pi}{N}lq }
\right\} ,
\label{discreteWF a} 
\nonumber \\ \\
\tilde{W}_{\hat{A}}(k,b)
&=& {\rm Tr}\left\{\hat{A}\left[\left(\hat{X}\hat{Z}^b \right)^k \right]^{\dagger} \right\}     \; ,
\label{discreteWF b}\\
\tilde{W}_{\hat{A}}(l)
&=& {\rm Tr}\left[\hat{A} \left(\hat{Z}^l\right)^{\dagger} \right] \; .
\label{discreteWF c}
\end{eqnarray}
\label{discreteWF}
\end{subequations}
The variables $q,p = 0,1, \cdots, N-1$ denote the coordinate and momentum in our discrete phase space, which thus consists of an $N \times N$ set of points.
The quantities $\hat{Z}$ and $\hat{X}$ appearing in Eqs. (\ref{discreteWF}) are the Schwinger operators, defined, for convenience, in 
App. \ref{schwinger}.
Definition (\ref{discreteWF}) is, for the discrete case, analogous to
that of Eqs. (\ref{WF continuous}) for the continuous case.
The $N(N-1)$ operators $\left(\hat{X}\hat{Z}^b \right)^k$ 
($b=0,1, \cdots, N-1; \; k=1, \cdots, N-1$) appearing in 
Eq. (\ref{discreteWF b}), together with the $N$ operators $\hat{Z}^l$
($l=0, \cdots, N-1$) appearing in Eq. (\ref{discreteWF c}), form a complete set of $N^2$ operators (see Eqs. (\ref{(XZb)m,Zl})).

We shall take the dimensionality $N$ to be a prime number larger than 2, as in this case the integers $0,1, \cdots, N-1$ constitute a mathematical field, with addition, subtraction, multiplication and division defined ${\rm Mod}N$
(see, e.g., Refs. \cite{wootters87,micha-WF012}).
This field plays a role similar to that of the real numbers in the continuous case studied in the previous section.
The quantity $\omega = {\rm exp}(2\pi i/N)$, one of the $N$-th roots of 1, will appear frequently in our analysis;
we shall agree that the numerical exponents of $\omega$ to be considered in what follows always belong to the ${\rm Mod}N$ algebra.
When the dimensionality $N$ is a prime number, we also know that the problem admits
exactly $N+1$ {\em mutually unbiased bases} (MUB) 
(see, e.g., Refs. \cite{micha-WF012,durt_et_al}).
The operators $\hat{X}\hat{Z}^b$, $b=0, \cdots N-1$ define $N$ of the $N+1$ MUB, [see Eq. (\ref{MUB})], while the operator $\hat{Z}$ defines the so-called ``reference basis", or ``computational basis".

\begin{figure}[h]
\centerline{\includegraphics[width=10cm]{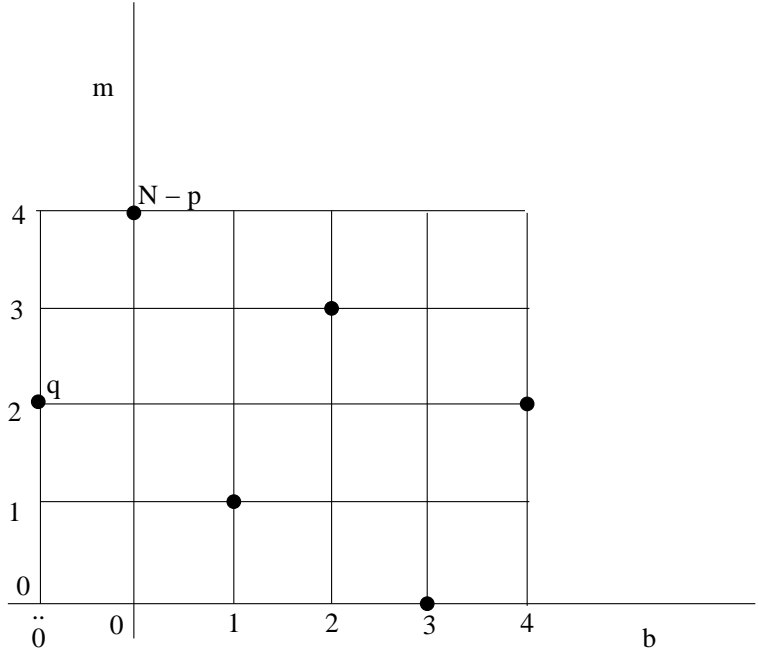}}
\caption{
\footnotesize{
Illustration of the function $m=M_{q,p}(b)$ in the $b-m$ plane, for $N=5$ and the particular pair of ``phase-space" values $q=2$, $p=1$. 
}}
\label{bm_plane}
\end{figure}

It is shown in App. \ref{wigner_discr_mub} that the definition (\ref{discreteWF}) can be written in terms of MUB as
\begin{subequations}
\begin{eqnarray}
W_{\hat{A}}(q,p)
&=&\frac{1}{N}
\sum_{b=\ddot{0}}^{N-1} \sum_{k=0}^{N-1} \sum_{m=0}^{N-1}
e^{\frac{2\pi i}{N}k \big[M_{q,p}(b)-m \big]}
\left\langle m; b\left|\hat{A}\right|m;b\right\rangle 
- {\rm Tr} (\hat{A})  \; ,
\nonumber \\
\label{discreteWF 1 a} 
\end{eqnarray}
where the reference basis has been denoted, for convenience, as $\ddot{0}$.
We have defined the quantity
\begin{eqnarray}
M_{q,p}(b) = 
\left\{
\begin{array}{cl}
(-p+bq) \; {\rm Mod}[N],  & {\rm for} \;\;\; b = 0, \cdots, N-1 \; , \\
q, & {\rm for} \;\;\; b=\ddot{0} \; .
\end{array}
\right.
\label{m(b)}
\end{eqnarray}
\label{discreteWF 1}
\end{subequations}
For a given pair of variables $q,p$, Eq. (\ref{m(b)}) states that, for $b=\ddot{0}$, $M_{q,p}(\ddot{0})=q$; 
for $b=0$, $M_{q,p}(0)=-p \; {\rm Mod}[N]=N-p$; 
for subsequent values of $b$, 
$M_{q,p}(b)=(-p+bq) \; {\rm mod}[N]$.
Thus, $M_{q,p}(b)$ may be viewed as specifying ``points" in a $b-m$ plane:
$b$ is along the $x$-axis and takes the values $b=\ddot{0},0,1,\cdots,N-1$, which denote the $N+1$ bases;
$m$ is along the $y$-axis and takes the values $b=0,1,\cdots,N-1$, which denote the $N$ states for each basis.
This aggregate of points, for {\em fixed} $q$ and $p$, may be described as a ``line" in the $b-m$ plane.
This is illustrated, for a particular case, in Fig. \ref{bm_plane}.
Further study, based on such a view, is in progress. 
We thus refer to $M_{q,p}(b)$ as a line, and its corresponding operator, $\hat{P}_{q,p}$, Eq. (\ref{Pj}) below, as a line operator;
it is similar to the ``phase-point" operator introduced in Secs. V and VI of 
Ref. \cite{wootters87}.

In Eq. (\ref{discreteWF 1 a}) we can do the sum over $k$, using the result
\begin{equation}
\frac{1}{N}\sum_{k=0}^{N-1}
e^{\frac{2\pi i}{N}k \big[M_{q,p}(b)-m \big]}
= \delta_{m,\; M_{q,p}(b)} \; ,
\label{sum_k}
\end{equation}
where the arguments of the Kronecker delta are understood to be ${\rm Mod}[N]$;
in other words, for given $q,p$, the sum (\ref{sum_k}) vanishes unless $m$ equals 
$(-p+bq) \; {\rm Mod}[N]$ when 
$b \neq \ddot{0}$, or $q$ when $b = \ddot{0}$.
Eq. (\ref{discreteWF 1 a}) can then be given the alternative forms
\begin{subequations}
\begin{eqnarray}
W_{\hat{A}}(q,p)
&=& \sum_{b=\ddot{0}}^{N-1} 
\left\langle M_{q,p}(b);b\left|\hat{A}\right|M_{q,p}(b);b \right\rangle 
- {\rm Tr} (\hat{A})  \; ,
\label{discreteWF 2 a} \\
&=& {\rm Tr} (\hat{A}\hat{P}_{q,p}) \; ,
\label{discreteWF 2 b}
\end{eqnarray}
where we have defined the Hermitean operator
\begin{eqnarray}
\hat{P}_{q,p}
=\sum_{b=\ddot{0}}^{N-1} 
\big|M_{q,p}(b);b\big\rangle  \big\langle M_{q,p}(b); b \big|
-\hat{\mathbb{I}},
\label{Pj}
\end{eqnarray}
\label{discrete_WF and line_operator}
\end{subequations}
$\hat{\mathbb{I}}$ being the unit operator.
From Eq. (\ref{discreteWF 1 a}), the line operator $\hat{P}_{q,p}$ can also be written more explicitly as
\begin{eqnarray}
\hat{P}_{q,p}
&=&\frac1N
\sum_{b=0}^{N-1}\sum_{k=1}^{N-1}\sum_{m=0}^{N-1}
e^{\frac{2 \pi i}{N}k(-p+bq-m)}
|m;b\rangle  \langle m;b| 
\nonumber \\
&&\hspace{2cm}+\frac1N \sum_{k=0}^{N-1}\sum_{n=0}^{N-1}
e^{\frac{2 \pi i}{N}k(q-n)} |n\rangle  \langle n|.
\label{line_operator 2}
\end{eqnarray}
Eqs. (\ref{line_operator 2}) and (\ref{Pj}) are analogous to 
Eq. (\ref{P(q,p) continuous mub a}) and (\ref{P(q,p) continuous mub b}), which correspond to the continuous case.
The integrals over $\theta$, $x'$ and $|t|$ of the continuous case correspond to the sums over $b$, $m$ and $k$ of the discrete one.

The WT of Eq. (\ref{discreteWF}) and the line operator of Eq. (\ref{Pj}) have the following properties, analogous to the ones for the continuous case.

1) As shown in App. \ref{matrix elements of Pqp}, the matrix elements of the line operator with respect to the states of the reference basis are given by
\begin{equation}
\langle q | \hat{P}_{q' p'} | \bar{q} \rangle
=\delta_{q q'} \delta_{\bar{q}q'}
-\delta_{q\bar{q}} \; \delta_{2q, \; 2q'+1}
+ \delta_{q + \bar{q},\; 2q'+1} \; e^{\frac{2\pi i}{N}p'(q-\bar{q})} \; .
\label{m. els. Pqp}
\end{equation}

2) The WT of a Hermitean operator $\hat{A}$ is real, i.e.,
\begin{equation}
W_{\hat{A}}(q,p) = W_{\hat{A}}^{\star}(q,p)\;, \;\;\;{\rm for}\;\;\; A^{\dagger}=A.
\label{W=W*}
\end{equation}
This follows immediately from the Hermiticity of the line operators $\hat{P}_{q,p}$.

3) The line operators $\hat{P}_{q,p}$ , $N^2$ in number, form a complete orthonormal set of operators, in the following sense:

i) It is shown in App. \ref{orthogonality Pj proof} that they fulfill the orthogonality relation
\begin{equation}
\frac{1}{N} {\rm Tr}\left[ \hat{P}_{q,p} \; \hat{P}_{q',p'} \right]
= \delta_{q, q'}\delta_{p, p'},
\label{orthogon. Pj}
\end{equation}
which is the discrete version of Eq. (\ref{orthog Ps cont}).

ii) From the expression (\ref{line_operator 2}), or from 
Eq. (\ref{Pj}), one finds, directly, that they satisfy the closure relation
\begin{equation}
\frac{1}{N} \sum_{q,p=0}^{N-1} \hat{P}_{q,p} = \mathbb{I},
\label{}
\end{equation}
which is the discrete version of Eq. (\ref{closure cont.}).

iii) An $N \times N$ matrix $\hat{A}$ can thus be written as a linear combination of the $\hat{P}_{q,p}$'s, i.e.,
\begin{subequations}
\begin{eqnarray}
\hat{A}
&=& \frac{1}{N} \sum_{q,p=0}^{N-1} 
{\rm Tr}\left(\hat{A} \hat{P}_{q,p}\right)\hat{P}_{q,p}
\label{A as lc of Pj a}   \\
&=& \frac{1}{N} \sum_{q,p=0}^{N-1} 
 W_{\hat{A}}(q,p) \hat{P}_{q,p} \; .
\label{A as lc of Pj b}
\end{eqnarray}
\label{A as lc of Pj}
\end{subequations}

4) The WT's of the operators $\hat{A}$ and $\hat{B}$ fulfill the so-called ``product formula" (see also Ref. \cite{wootters87}, Eq. (15))
\begin{equation}
\frac{1}{N}
\sum_{q,p=0}^{N-1} W_{\hat{A}}(q,p) W_{\hat{B}}(q,p)
= {\rm Tr}(\hat{A} \hat{B}) \; ,
\label{inner-product-rule}
\end{equation}
in analogy with Eq. (\ref{product formula continuous}) for the continuous case.
This can be proved as follows. 
From Eq. (\ref{A as lc of Pj b}) applied  to the operators $\hat{A}$ and $\hat{B}$, and using the orthogonality relation (\ref{orthogon. Pj}), we have
\begin{subequations}
\begin{eqnarray}
{\rm Tr}(\hat{A} \hat{B})
&=& \frac{1}{N^2} \sum_{q,p=0}^{N-1} \sum_{q',p'=0}^{N-1}
W_{\hat{A}}(q,p) W_{\hat{B}}(q',p')
{\rm Tr}\left[ \hat{P}_{q,p} \; \hat{P}_{q',p'} \right]  
\label{product formula a}   \\
&=& \frac{1}{N}
\sum_{q,p=0}^{N-1} W_{\hat{A}}(q,p) W_{\hat{B}}(q,p).
\label{product formula b}
\end{eqnarray}
\label{product formula}
\end{subequations}

5) The WF $W_{\hat{\rho}}(q,p)$ satisfies the marginality property, written in terms of the projector 
$\hat{\mathbb{P}}_{mb} = |m,b\rangle \langle m,b |$,
\begin{equation}
{\rm Tr}\left(\hat{\rho} \; \hat{\mathbb{P}}_{mb} \right)
=\langle m,b| \hat{\rho} |m,b\rangle
= \frac{1}{N}\sum_{q,p=0}^{N-1}W_{\hat{\rho}}(q,p) \delta_{M_{q,p}(b),m} \; ,
\label{marginality disc}
\end{equation}
where we recall that $M_{q,p}(b)$ is defined in Eq. (\ref{m(b)}).

Eq. (\ref{marginality disc}), analogous to (\ref{marginality cont}) for the continuous case, states that the probability to find the system 
in the state $m$ of the basis $b$ (of our set of $N+1$ MUBs) is $1/N$ times the sum of the WF over the points in the phase-space plane $q,p$ that satisfy 
$M_{q,p}(b)=m$.
The marginality relation, Eq. (\ref{marginality disc}), is obtained from the product formula
(\ref{product formula b}) for $\hat{A}=\hat{\rho}$ and $B=\hat{\mathbb{P}}_{mb}$, the WT of the latter being, from 
Eq. (\ref{discreteWF 2 b})
\begin{subequations}
\begin{eqnarray}
W_{\hat{\mathbb{P}}_{mb}}(q,p)
&=& {\rm Tr} (\hat{\mathbb{P}}_{mb} \hat{P}_{qp})
\\
&=& {\rm Tr} \left\{
|m;b\rangle \langle m;b|
\left[
\sum_{b'=\ddot{0}}^{N-1} 
\left|M_{q,p}(b');b'\right\rangle  \left\langle M_{q,p}(b'); b' \right|
-\hat{\mathbb{I}}
\right]
\right\}
\label{WT Pmb a} 
\nonumber \\ \\
&=&  
\big|\big\langle m;b \big|M_{q,p}(b);b \big\rangle \big|^2
+\sum_{b'=\ddot{0}, (\neq b)}^{N-1} 
\big|\big\langle m;b \big|M_{q,p}(b');b' \big\rangle\big|^2
-1
\label{WT Pmb b}  \\
&=& \delta_{M_{q,p}(b),m} + N\frac{1}{N} - 1 
\label{WT Pmb c}  \\
&=& \delta_{M_{q,p}(b),m}  .
\end{eqnarray}
\label{WT Pmb}
\end{subequations}
We comment in passing that the RHS of Eq. (\ref{marginality disc})
can be considered as defining the {\em Radon transform} 
of the WF $W_{\hat{\rho}}(q,p)$ 
(see, e.g., Refs. \cite{leonhardt,schleich,khanna-mello-revzen,micha-WF012}).

Two particular cases of the above marginality property are: 
i) $b=\ddot{0}$: then $m$ is a coordinate, which we may call $q_0$, 
and the resulting summation in phase space
(i.e., the RHS of (\ref{marginality disc}); see Eq. (\ref{m(b)})) is over the {\em line in the $(q,p)$-plane} containing all $p$'s for that $q_0$; i.e.,
\begin{equation}
b=\ddot{0}:  \hspace{5mm} 
{\ Tr}(\hat{\rho}\hat{\mathbb{P}}_{q_0,\ddot{0}})
=\langle q_0| \hat{\rho} |q_0\rangle
= \frac{1}{N}\sum_{q,p}W_{\hat{\rho}}(q,p) \delta_{q,q_0}
=\frac{1}{N}\sum_{p}W_{\hat{\rho}}(q_0,p) \; ;
\label{marginal q0} 
\end{equation}
ii) for $b=0$, we identify $m=N-p_0$ (see Eq. (\ref{b=0 states c})),
and the resulting summation in phase space 
(i.e., the RHS of Eq. (\ref{marginality disc}); see Eq. (\ref{m(b)}):
$M_{q,p}(0)=N-p$)
is over the {\em line in the $(q,p)$-plane} containing all $q$'s for that $p_0$; i.e.,
\begin{subequations}
\begin{eqnarray}
b=0: \hspace{5mm} 
{\ Tr}(\hat{\rho}\hat{\mathbb{P}}_{N-p_0,0})
&=& \frac{1}{N}\sum_{q,p}W_{\hat{\rho}}(q,p) \delta_{N-p,N-p_0}, \\
{\rm i.e.,} \;\;\; \langle p_0| \hat{\rho} |p_0\rangle
&=&\frac{1}{N}\sum_{q}W_{\hat{\rho}}(q,p_0)
\end{eqnarray}
\end{subequations}
These are the standard marginality relations, which can also be obtained trivially from the form 
(\ref{discrete_WF and line_operator}) for the WF, without using the product formula.
For the case $b=1,\cdots,N$, Eq. (\ref{marginality disc}) states that
\begin{equation}
\langle m,b| \hat{\rho} |m,b\rangle
= \frac{1}{N}\sum_{q,p}W_{\hat{\rho}}(q,p) \delta_{-p+bq,m} \; ,
\label{marginal 2}
\end{equation}
the sum on the RHS being over the points on the
{\em line in phase space $(q,p)$} defined by $-p+bq=m {\rm Mod}[N]$, for fixed $m,b$.

6) The WF is normalized as 
\begin{equation}
\frac{1}{N}
\sum_{p,q=0}^{N-1} W_{\hat{\rho}}(q,p)
=1 \; .
\label{normal}
\end{equation}
just as in Eq. (\ref{normalization}) for the continuous case.

Notice that the various properties mentioned in the previous section for the continuous case can be translated to the discrete case with the correspondence 
$1/2\pi \Rightarrow 1/N \;$. 

\subsection{Relation between Wigner function and Kirkwood quasi-distribution for a discrete, finite-dimensional hilbert space}
\label{}

Going back to our program of relating WF to Kirkwwod's quasi-distribution,
we show in App. \ref{derivation of W(K) discrete} the relation
\begin{eqnarray}
W_{\hat{\rho}}(q,p)
&=& \sum_{q',p'=0}^{N-1} e^{\frac{2\pi i}{N}2 (q-q'+\frac{N+1}{2})(p-p')}
K_{p'q'} 
\nonumber \\
&&\hspace{1cm} + \langle  q|\hat{\rho}| q \rangle
- \Big\langle q + (N+1)/2 \Big| \hat{\rho} 
\Big| q + (N+1)/2 \Big\rangle .
\label{W(K) discrete 1}
\end{eqnarray}
Notice that in this equation the labels occurring in bras and kets must
be understood ${\rm Mod}[N]$.
It can be checked directly that the result (\ref{W(K) discrete 1}) fulfills the normalization condition (\ref{normal}).

The result of Eq. (\ref{W(K) discrete 1}) is analogous to that of Eqs. (\ref{W(K) continuous 1}) and 
(\ref{W(K) 2}) for the continuous case. 
The Kirkwood distribution $K(p,q)$ is defined as in 
Eqs. (\ref{Kpq, Pq, Pp continuous}) for the continuous case, except that the states $|q\rangle$ and $|p)$ are to be defined as in Eqs. (\ref{Z}) and 
(\ref{ket p (kets q)}).

Just as in the previous section, we notice from  Eq. (\ref{K(pm,qn) 1}) that Kirkwood's distribution can be related to the correlations of two probes, in a very weak-coupling measurement designed to measure in succession the projectors for position and momentum of the system. 
For the present discrete case, ($p,q=0,\cdots,N-1$), we write relation
(\ref{K(pm,qn) 1}) as
\begin{equation}
K(p,q)
= {\rm lim}_{\epsilon_1 \to 0} \frac{1}{\epsilon_1 \epsilon_2}
\left[\langle \hat{Q}_1 \hat{Q}_2 \rangle^
{(\mathbb{P}_{p} \leftarrow \mathbb{P}_{q})} 
+ \frac{i}{2 \sigma_{P_1}^2} 
\langle \hat{P}_1 \hat{Q}_2 \rangle
^{(\mathbb{P}_{p} \leftarrow \mathbb{P}_{q})} 
\right].
\label{Kpq discrete}
\end{equation}
Substituting this relation in Eq. (\ref{W(K) discrete 1}) we thus find
\begin{eqnarray}
W_{\hat{\rho}}(q,p)
&=& \sum_{q',p'=0}^{N-1} e^{\frac{2\pi i}{N}2(q-q'+(N+1)/2)(p-p')}
\nonumber \\
&& \hspace{1cm} \times\left\{
{\rm lim}_{\epsilon_1 \to 0} \frac{1}{\epsilon_1 \epsilon_2}
\left[\langle \hat{Q}_1 \hat{Q}_2 \rangle^
{(\mathbb{P}_{p'} \leftarrow \mathbb{P}_{q'})} 
+ \frac{i}{2 \sigma_{P_1}^2} 
\langle \hat{P}_1 \hat{Q}_2 \rangle
^{(\mathbb{P}_{p'} \leftarrow \mathbb{P}_{q'})} 
\right]
\right\} 
\nonumber \\
&&
+\frac{1}{\epsilon}
\langle \hat{Q} \rangle^{(\mathbb{P}_q)}
-\frac{1}{\epsilon}
\langle \hat{Q} \rangle^{(\mathbb{P}_{q+(N+1)/2})} \; .
\label{W(probe correl)discrete}
\end{eqnarray}
The last two terms in (\ref{W(probe correl)discrete}) are the expectation value of the probe position in a single measurement designed to pre-measure the projectors
$\mathbb{P}_q$ and $\mathbb{P}_{q+(N+1)/2}$, respectively.

Result (\ref{W(probe correl)discrete}) is the discrete Hilbert-space counterpart of the duly discretized continuous case that was given in
Eq. (\ref{W(probe correl)continuous}) of the previous section.

As a result, Wigner function, which is defined in the system discrete phase space, can be related to a set of {\em measurable quantities}, consisting of the two-probe and single-probe expectation values obtained in the experimental setup described above, and reconstructed therefrom.

\section{Conclusions}
\label{concl}

In this paper we posed the question whether it is possible to find appropriate measurements involving the system position and momentum that would allow the reconstruction of Wigner function of the system state.
We were able to give an affirmative answer to this question.
The type of measurements needed are generalizations of the model 
envisaged by von Neumann in his model of QM measurement.
They involve successive couplings of two probes with projectors associated with the system position and momentum.
In this model, what one detects are the correlation functions of the two probes, which are compatible dynamical variables, not the system itself.

We first considered the case in which the system is described in a continuous Hilbert space, and then we turned to the study of a description in a discrete, finite-dimensional Hilbert space.

The Wigner function for this latter case of a discrete, finite-dimensional Hilbert space, has been widely studied in the literature.
Here we proposed an alternative version, formulated, in this paper, within a standard algebraic approach; however, as it turns out, this version can be re-formulated entirely in terms of ``finite-geometry" concepts, an approach that associates states and operators in Hilbert space with lines and points of the geometry \cite{micha-WF012}.
This latter approach is conceptually very attractive, and its development will be postponed to a future publication.

\section{Acknowledgments}
One of the authors (PAM) acknowledges support from the Sistema Nacional de Investigadores (M\'exico), and Conacyt grant No. 79501.
He is also grateful to the Physics Department of the Technion, Haifa, for its
hospitality during a stay where part of this work was developed.
MR acknowleges gratefully informative discussions with Prof. A. Mann.

\appendix

\section{Derivation of the relation Eq. (\ref{W(K) continuous 1}) between WF and Kirkwood quasi-distribution for the continuous case}
\label{derivation of W(K) continuous}

Wigner function is defined in Eq. (\ref{WF continuous}).
Using the BCH identity (Ref. \cite{peres}, p. 333)
\begin{equation}
e^{\hat{A}+\hat{B}}
= e^{\hat{A}} e^{\hat{B}} e^{-\frac12 [\hat{A},\hat{B}]} ,
\label{BCH}
\end{equation}
valid when $\hat{A}$ and $\hat{B}$ commute with their commutator, and expressing 
the operators $e^{-iv\hat{p}}$ and 
$e^{-iu\hat{q}}$ in their spectral representation, we can write
$\tilde{W}_{\hat{\rho}}(u,v)$ as 
\begin{equation}
\tilde{W}_{\hat{\rho}}(u,v)
= e^{-\frac{i}{2}uv}
\int_{-\infty}^{\infty} \int_{-\infty}^{\infty}
dq' dp' e^{-i(u q' + v p')}
K(p',q') \; ,
\label{FTrho 1}
\end{equation}
where $K(q,p)$ is Kirkwood's quasi-distribution \cite{kirkwood,dirac} 
of Eq. (\ref{K continuous 1}).
Introducing (\ref{FTrho 1}) in (\ref{WF continuous a}) and using the result
\begin{equation}
\frac{1}{4 \pi} \int_{-\infty}^{\infty}\int_{-\infty}^{\infty}
e^{\frac{i}{2}(\xi u + \eta v - uv)} du dv
= e^{\frac{i}{2}\xi \eta} \; ,
\label{}
\end{equation}
we find Eq. (\ref{W(K) continuous 1}).

An alternative derivation of this result is based on the standard definition of WT of an operator $\hat{A}$ (see, e.g., 
Refs. \cite{wigner,leonhardt,schleich,khanna-mello-revzen})
\begin{equation}
W_{\hat{A}}(q,p)
= \int_{-\infty}^{\infty}e^{-ipy}
\left\langle q+\frac{y}{2} \right|\hat{A} \left| q+\frac{y}{2} \right\rangle.
\label{WT,continuous,standard}
\end{equation}

The Kirkwood distribution of Eq. (\ref{K continuous 1}) can be written in terms of Wigner functions using the so-called ``overlap formula", or ``product formula", as
\begin{equation}
K(p,q)
= \int \int W_{\hat{\rho}}(q',p')W_{\hat{\mathbb{P}}_p \hat{\mathbb{P}}_q}(q',p')
\frac{dq'dp'}{2\pi}.
\label{K(WW)}
\end{equation}
We have
\begin{subequations}
\begin{eqnarray}
W_{\hat{\mathbb{P}}_p \hat{\mathbb{P}}_q}(q',p')
&=& \int e^{-ip'y} \left\langle q'+\frac{y}{2}\right|
\hat{\mathbb{P}}_p \hat{\mathbb{P}}_q \left|q'-\frac{y}{2}\right\rangle dy 
\label{W_PP a} \\
&=& \frac{1}{\pi} e^{-2i(p-p')(q-q')}
\label{W_PP b}
\end{eqnarray}
\label{W_PP}
\end{subequations}
From (\ref{K(WW)}) we find
\begin{subequations}
\begin{eqnarray}
&& \int \int
K(\bar{p}, \bar{q})
e^{2i(p-\bar{p})(q-\bar{q})}
d\bar{q}d\bar{p}
\nonumber \\
&& \hspace{1cm} = \int \int \int \int 
W_{\rho}(q',p')W_{\mathbb{P}_{\bar{p}} \mathbb{P}_{\bar{q}}}(q',p')
e^{2i(p-\bar{p})(q-\bar{q})}
\frac{dq'dp'}{2\pi} d\bar{q} d\bar{p}
\nonumber \\
\label{K(WW) 1a} \\
&& \hspace{1cm} = \int \int \int \int 
W_{\rho}(q',p')
\left[
\frac{1}{\pi} e^{-2i(\bar{p}-p')(\bar{q}-q')}
\right]
e^{2i(\bar{p}-p)(\bar{q}-q)}
\frac{dq'dp'}{2\pi} d\bar{q} d\bar{p} \; ,
\label{K(WW) 1b} \nonumber \\ \\
&&  \hspace{10mm}= \frac12 W_{\rho}(p,q) \; ,
\label{K(WW) 1c}
\end{eqnarray}
\label{K(WW) 1}
\end{subequations}
which is the result (\ref{W(K) continuous 1}).
From (\ref{K(WW) 1a}) to  (\ref{K(WW) 1b}) we have used the result 
(\ref{W_PP b}).
From (\ref{K(WW) 1b}) to (\ref{K(WW) 1c}) we have used the identity
\begin{equation}
\int \int
e^{-2i(\bar{p}-p')(\bar{q}-q')}
e^{2i(\bar{p}-p)(\bar{q}-q)}
d\bar{q} d\bar{p}
= \pi^2  \delta(p'-p) \delta(q'-q) \; .
\label{}
\end{equation}

\section{von Neumann model for position and momentum}
\label{vnM x,p}

The operator
$
\hat{\mathbb{P}}_q = |q \rangle \langle q|
\label{Pq}
$
of Eq. (\ref{Pq, continuous}) is not a proper position projector, since it is not idempotent.
In order to use the formalism developed in Ref. \cite{lars-mello}, we define the operators (with $q_n=n\delta$)
\begin{equation}
\mathbb{P}_{q_n} 
= \int_{q_n -\delta/2}^{q_n +\delta/2}|q \rangle dq \langle q|,
\label{Pxn}
\end{equation}
which have the properties
\begin{subequations}
\begin{eqnarray}
\mathbb{P}_{q_n} \mathbb{P}_{q_{n'}} 
&=& \delta_{nn'} \mathbb{P}_{q_n}
\label{PnPm} \\
\sum_{n=-\infty}^{\infty} \mathbb{P}_{q_n} &=& 1.
\label{closure}
\end{eqnarray}
\end{subequations}
Similarly, for the momentum we define the operators
\begin{equation}
\mathbb{P}_{p_m} 
= \int_{p_m -\delta/2}^{p_m +\delta/2}|p \rangle dp \langle p|,
\label{Ppm}
\end{equation}
which have similar properties.

The approximation involved in going from Eq. (\ref{W(K) 2 a}) to (\ref{W(K) 2 b}) in the text can be justified in terms of $c$-number functions in the following way.
For any two $\langle \phi|$ and $|\psi\rangle$, consider the following integral and the approximations to it given in the subsequent equations
\begin{subequations}
\begin{eqnarray}
\langle \phi|     
\int_{-\delta/2}^{\delta/2}f(q') \mathbb{P}_{q'} dq' |\psi\rangle
&=&
\int_{-\delta/2}^{\delta/2}f(q') \phi^{*}(q') \psi(q')dq'   
\label{discrete approximation a}  \\
&\approx& f(q_1)\int_{-\delta/2}^{\delta/2}\phi^{*}(q') \psi(q')dq' 
\label{discrete approximation b}  \\
&=& \langle \phi|f(q_1) \mathbb{P}_{q_0} |\psi\rangle ,
\label{discrete approximation c}
\end{eqnarray}
\label{discrete approximation}
\end{subequations}
for some suitable $q_1 \in [-\delta/2, \delta/2]$. 
In (\ref{discrete approximation c}) we have used the notation of Eq. (\ref{Pxn}).
With this argument we thus approximate the operator inside the first bracket in 
(\ref{discrete approximation a}) by the one in (\ref{discrete approximation c}).

Ref. \cite{lars-mello} studies the extension to two probes of von Neumman's measurement model (vNM)
\begin{equation}
\hat{H}(t)
= \epsilon_1 \delta(t-t_1) \hat{\mathbb{P}}_{q_n} \hat{P}_1
+\epsilon_2 \delta(t-t_2) \hat{\mathbb{P}}_{p_m} \hat{P}_2, 
\hspace{1cm} 0<t_1<t_2
\label{vNM}
\end{equation}
in which $\hat{\mathbb{P}}_{q_n}$ plays the role of the observable to be pre-measured first and $\mathbb{P}_{p_m}$, later.

The position-position and momentum-position correlation of the two probes is found to be
\begin{subequations}
\begin{eqnarray}
\frac{1}{\epsilon_1 \epsilon_2}
\langle \hat{Q}_1 \hat{Q}_2 \rangle^{(\hat{\mathbb{P}}_{p_m} 
\leftarrow \hat{\mathbb{P}}_{q_n})}
&=& \Re W_{11}^{(\hat{\mathbb{P}}_{p_m} \leftarrow \hat{\mathbb{P}}_{q_n})}
(\epsilon_1)
\label{Q1Q2} \\
\frac{1}{\epsilon_1 \epsilon_2}
\langle \hat{P}_1 \hat{Q}_2 \rangle^{(\hat{\mathbb{P}}_{p_m} \leftarrow \hat{\mathbb{P}}_{q_n})}
&=& 2 \sigma_{P_1}^2
\Im W_{11} ^{(\hat{\mathbb{P}}_{p_m} \leftarrow \hat{\mathbb{P}}_{q_n})}
(\epsilon_1) ,
\label{Q1Q2}
\end{eqnarray}
\label{Q1Q2,P1Q2}
\end{subequations}
where
\begin{subequations}
\begin{eqnarray}
W_{11}^{(\hat{\mathbb{P}}_{p_m} \leftarrow \hat{\mathbb{P}}_{q_n})}
(\epsilon_1)
&=& \sum_{n'} G_{n' n}(\epsilon_1) 
{\rm Tr}_s 
\left(\hat{\rho}_s \hat{\mathbb{P}}_{q_{n'}} \hat{\mathbb{P}}_{p_m} \hat{\mathbb{P}}_{q_n}\right)
\label{W11}   \\
G_{n' n}(\epsilon_1)
&=& \delta_{n n'} + e^{-\frac12 \sigma_{P_1}^2 \epsilon_1^2}(1-\delta_{n n'}).
\label{Gnn'}
\end{eqnarray}
\label{W11,Gnn'}
\end{subequations}
A Gaussian distribution for the original state of the probes is assumed, and
$\sigma_{P_1}^2$ denotes the momentum variance of probe 1.
In the limit $\epsilon_1 \to 0$, the above expression (\ref{W11}) becomes
\begin{equation}
W_{11}^{(\hat{\mathbb{P}}_{p_m} \leftarrow \hat{\mathbb{P}}_{q_n})}(0)
= {\rm Tr}_s 
\left(\hat{\rho}_s \hat{\mathbb{P}}_{p_m} \hat{\mathbb{P}}_{q_n}\right)
\equiv K(p_m,q_n),
\label{K(pm,qn)}
\end{equation}
which is Kirkwood's joint quasi-probability distribution \cite{kirkwood,dirac} for the variables $p_m$ and $q_n$, in the original state of the system $\rho_s$.

Using Eqs. (\ref{Q1Q2,P1Q2}), Kirkwood's joint quasi-distribution can thus be expressed in terms of measurements performed on the probes as
\begin{equation}
K(p_m,q_n)
= {\rm lim}_{\epsilon_1 \to 0} \frac{1}{\epsilon_1 \epsilon_2}
\left[\langle \hat{Q}_1 \hat{Q}_2 \rangle^
{(\hat{\mathbb{P}}_{p_m} \leftarrow \hat{\mathbb{P}}_{q_n})} 
+ \frac{i}{2 \sigma_{P_1}^2} 
\langle \hat{P}_1 \hat{Q}_2 \rangle
^{(\hat{\mathbb{P}}_{p_m} \leftarrow \hat{\mathbb{P}}_{q_n})} 
\right].
\label{K(pm,qn) 1}
\end{equation}

\section{Schwinger operators and MUB}
\label{schwinger}

We consider our $N$-dimensional Hilbert space to be spanned by $N$ distinct states 
$|q\rangle$, with $q=0,1, \cdots ,(N-1)$, which are subject to the periodic condition
$|q+N\rangle=|q\rangle$.
These states are designated as the ``reference basis", or ``computational basis" of the space.  
We shall follow Schwinger \cite{schwinger} and introduce the unitary operators $\hat{X}$ and $\hat{Z}$, defined by their action on the states of the reference basis by the equations
\begin{subequations}
\begin{eqnarray}
\hat{Z}|q\rangle
&=&\omega^q|q\rangle, \;\;\;\; \omega=e^{2 \pi i/N},
\label{Z}  \\
\hat{X}|q\rangle &=& |q+1\rangle .
\label{X}
\end{eqnarray}
\label{Z,X}
\end{subequations}
The operators $\hat{X}$ and $\hat{Z}$ fulfill the periodic condition
\begin{equation}
\hat{X}^N = \hat{Z}^N = 
\hat{\mathbb{I}},
\label{X,Z periodic}
\end{equation}
$\hat{\mathbb{I}}$ being the unit operator.
These definitions lead to the commutation relation
\begin{equation}
\hat{Z}\hat{X}=\omega \hat{X}\hat{Z} .
\label{comm Z,X}
\end{equation}
The two operators $\hat{Z}$ and $\hat{X}$ form a complete algebraic set, in that only a multiple of the identity commutes with both \cite{schwinger}. 
As a consequence, any operator defined in our $N$-dimensional Hilbert space can be written as a function of $\hat{Z}$ and $\hat{X}$.

We introduce the Hermitean operators  $\hat{p}$ and $\hat{q}$,
which play the role of ``momentum" and ``position", through the equations
\cite{de_la_torre-goyeneche,durt_et_al}
\begin{subequations}
\begin{eqnarray}
\hat{X}
&=& \omega^{-\hat{p}}
= e^{-\frac{2\pi i}{N}\hat{p}} \; ,
\label{X(p)} \\
\hat{Z}
&=& \omega^{\hat{q}}
=e^{\frac{2\pi i}{N}\hat{q}} \; .
\label{Z(q)} 
\end{eqnarray}
\label{X(p),Z(q)}
\end{subequations}
What we defined as the reference basis can thus be considered as the ``position basis".
With (\ref{comm Z,X}) and definitions (\ref{X(p),Z(q)}), the commutator of $\hat{q}$ and $\hat{p}$ in the continuous limit \cite{de_la_torre-goyeneche,durt_et_al}
is the standard one, $[\hat{q},\hat{p}]=i$.

The ``momentum basis" consists of the eigenstates of $\hat{X}$, which can be expanded in terms of the position basis as
\begin{subequations}
\begin{eqnarray}
|p \rangle
= \sum_{q=0}^{N-1}\frac{e^{\frac{2\pi i}{N}pq}}{\sqrt{N}} \; |q\rangle \; ,
\label{ket p (kets q)}
\end{eqnarray}
and satisfy the eigenvalue equation 
(see Ref. \cite{de_la_torre-goyeneche}, Eq. (12))
\begin{eqnarray}
\hat{X}|p\rangle = e^{-\frac{2\pi i}{N}p} |p\rangle.
\label{ev eqn X(p)}
\end{eqnarray}
\end{subequations}

The $N^2$-dimensional matrix space is spanned by the complete orthonormal $N^2$ operators $\hat{X}^m \hat{Z}^l$, with $m,l=0,1,..(N-1)$, so that any 
$N \times N$ matrix can be written as a linear combination of these $N^2$ operators.
A familiar example is a $2$-dimensional Hilbert space, where any 
$2\times 2$ matrix can be written as a linear combination of the three Pauli matrices plus the unit matrix, which can also be written as $\sigma_x$, $\sigma_z$, $\sigma_x \sigma_z$ and $I$.


For $N={\rm prime}\; >2$, we find the following identities:
\begin{subequations} 
\begin{eqnarray}
(\hat{X} \hat{Z}^{b})^k
&=&\omega ^{\frac{k(k-1)}{2}b} \hat{X}^k\hat{Z}^{kb}  
\label{(XZb)m vs Xm.Zmb}   \\
&=&\omega ^{-\frac{k(k+1)}{2}b}\hat{Z}^{kb} \hat{X}^k
\label{(XZb)m vs Zmb.Xm} \\
\hat{X}^k \hat{Z}^{l} 
&=&\omega^{-kl} \hat{Z}^{l} \hat{X}^k 
\label{XmZl vs ZlXm} \\
(\hat{X} \hat{Z}^{b})^N
&=&\hat{I}
\label{(XZb)N}
\end{eqnarray}
\label{(XZb)m,Xm.Zmb,Zmb.Xm}
\end{subequations}

Our complete orthonormal set of $N^2$ operators can be taken as
\begin{subequations} 
\begin{eqnarray}
&&(\hat{X} \hat{Z}^{b})^k, \hspace{1cm}b=0,1,\cdots,N-1, \\
&& \hspace{2.5cm} k=1, \cdots, N-1, \nonumber \\
&&\hat{Z}^l \; , \hspace{2cm}l=0,1,\cdots,N-1 \; .
\end{eqnarray}
\label{(XZb)m,Zl}
\end{subequations}

The operator $\hat{X} \hat{Z}^{b}$ possesses $N$ eigenvectors, denoted by 
$|m,b\rangle$ (see Eqs. (10), (11) of Ref. \cite{micha-WF012})
\begin{subequations} 
\begin{eqnarray}
\hat{X}\hat{Z}^{b} |m,b\rangle
&=& \omega^m |m;b\rangle ,
\label{e-value eqn mub} \\
|m;b\rangle
&=& \frac{1}{\sqrt{N}}\sum_{n=0}^{N-1}\omega^{\frac{b}{2}n(n-1)-nm} |n\rangle,
\;\;\;\; b,m = 0,1, \cdots , N-1.
\nonumber \\
\label{mub e-vectors}
\end{eqnarray}
\label{MUB}
\end{subequations}
Here, $|n\rangle$ ($n=0, \cdots, N-1$) denote the $N$ states of the reference basis. We have, altogether, $N+1$ mutually unbiased bases (MUB).
The states with $b=0$, i.e.,
\begin{subequations} 
\begin{eqnarray}
|m;0\rangle
&=& \frac{1}{\sqrt{N}}\sum_{n=0}^{N-1}
e^{-\frac{2\pi i}{N}mq} |q\rangle, 
\label{b=0 states a} \\
&=&\frac{1}{\sqrt{N}}\sum_{n=0}^{N-1}
e^{\frac{2\pi i}{N}(N-m)q} |q\rangle \; ,
\label{b=0 states b}
\end{eqnarray}
are eigenstates of $\hat{p}$ which, from Eq. (\ref{ket p (kets q)}), can be written as
\begin{eqnarray}
|m;0\rangle
&=& |p=-m=(N-m) {\rm Mod}[N]\rangle.
\label{b=0 states c}
\end{eqnarray}
\label{b=0 states}
\end{subequations}

\section{Derivation of Eqs. (\ref{discreteWF 1}) for the discrete Wigner Function}
\label{wigner_discr_mub}

We can write the quantities $\tilde{W}_{\hat{A},\hat{B}}(k,b)$, 
$\tilde{W}_{\hat{A},\hat{B}}(l)$ appearing in Eqs. (\ref{discreteWF}) in terms of the MUB basis $|m,b\rangle$ defined in Eqs. (\ref{MUB}).
The operator $\hat{X} \hat{Z}^{b}$ can be written in the spectral representation as
\begin{subequations}
\begin{eqnarray}
\hat{X} \hat{Z}^{b} 
&=& \sum_{m=0}^{N-1} |m, b\rangle \omega^m \langle m,b| \; ,
\label{XZb spectral} \\
\left[(\hat{X} \hat{Z}^{b})^k \right]^{\dagger}
&=& \sum_{m=0}^{N-1} |m, b\rangle \omega^{-mk} \langle m,b| \; ,
\label{(XZb)k spectral} 
\end{eqnarray}
\label{XZb (XZb)k spectral}
\end{subequations}
so that
\begin{equation}
\tilde{W}_{\hat{A}}(k,b) 
= \sum_{m=0}^{N-1}\omega^{-mk}
\langle m,b |\hat{A}| m,b \rangle \; ,
\label{WtildeA k,b 1}  
\end{equation}
Similarly,
\begin{subequations}
\begin{eqnarray}
\hat{Z} 
&=& \sum_{m=0}^{N-1} |n\rangle \omega^n \langle n| \; ,
\label{Z spectral} \\
(\hat{Z}^{l})^{\dagger}
&=& \sum_{n=0}^{N-1} |n\rangle \omega^{-nl} \langle n| \; ,
\label{Zl spectral} 
\end{eqnarray}
\label{Z, Zl spectral}
\end{subequations}
so that
\begin{equation}
\tilde{W}_{\hat{A}}(l) 
= \sum_{n=0}^{N-1}\omega^{-nl}
\langle n |\hat{A}| n \rangle \; .
\label{Wtilde A l 1} 
\end{equation}
Substituting these results in Eqs. (\ref{discreteWF}), we obtain 
Eqs. (\ref{discreteWF 1}).

\section{Derivation of the relation (\ref{m. els. Pqp}) for the matrix elements of the line operator}
\label{matrix elements of Pqp}

In the definition (\ref{Pj}) of the line operator we single out the first two terms, to write
\begin{equation}
\hat{P}_{q',p'}
= |q'\rangle \langle q'| + |N-p';0\rangle \langle N-p';0|
+ \sum_{b=1}^{N-1} 
\left|-p'+bq';b\right\rangle  \left\langle -p'+bq'; b \right|
-\hat{\mathbb{I}}.
\label{Pj 1}
\end{equation}
Recall, from Eqs. (\ref{b=0 states}), that 
$|N-p';0 \rangle = |p'\rangle$, and that $-p'+bq'$ is understood ${\rm Mod}N$.

Using Eq. (\ref{mub e-vectors}) for the states of the MUB, we write the matrix element $\langle q | \hat{P}_{q',p'}|\bar{q}\rangle$ as
\begin{subequations}
\begin{eqnarray}
\langle q | \hat{P}_{q',p'}|\bar{q}\rangle
&=& \delta_{qq'}\delta_{\bar{q}q'} 
+\frac1N e^{\frac{2\pi i}{N}p'(q-\bar{q})}
-\delta_{q,\bar{q}} 
\nonumber \\
&& +\frac1N \sum_{b=1}^{N-1} 
e^{\frac{2\pi i}{N}\left[\frac{b}{2} q(q-1)-q(-p'+bq')\right]}
e^{-\frac{2\pi i}{N}\left[\frac{b}{2} \bar{q}(\bar{q}-1)-\bar{q}(-p'+bq')\right]} 
\label{Pj 2 b} 
\nonumber \\ \\
&=& \delta_{qq'}\delta_{\bar{q}q'} 
+\frac1N e^{\frac{2\pi i}{N}p'(q-\bar{q})}
-\delta_{q,\bar{q}} 
+\frac1N (\alpha-1)e^{\frac{2\pi i}{N}p'(q-\bar{q})} \; .
\nonumber \\
\label{Pj 2 c}
\end{eqnarray}
\label{Pj 2}
\end{subequations}
The quantity $\alpha$ is defined as 
\begin{equation}
\alpha= \sum_{b=0}^{N-1} e^{\frac{2\pi i}{N} b \left[\frac12(q-\bar{q})(q+\bar{q}-1-2q')\right]} 
\label{X}
\end{equation}
and is nonzero only when 
\begin{equation}
\frac12(q-\bar{q})(q+\bar{q}-1-2q')
=0 \; {\rm Mod}[N] \; .
\label{Xneq0}
\end{equation}
I.e., 
\begin{equation}
\alpha = N\left[
\delta_{q,\;\bar{q}} + \delta_{q+\bar{q}, \; 1+2q'}
-\delta_{q,\; \bar{q}}\delta_{q+\bar{q}, \; 1+2q'}
\right],
\label{X 1}
\end{equation}
where the arguments of the Kronecker deltas are understood, as always, 
${\rm Mod}[N]$.
Substituting (\ref{X 1}) in Eq. (\ref{Pj 2 c}), we find the result 
(\ref{m. els. Pqp}).

\section{Proof of the orthogonality relation, Eq. (\ref{orthogon. Pj})}
\label{orthogonality Pj proof}

The definition of a ``line", Eq. (\ref{m(b)}), implies that two distinct lines, i.e., such that their parameters $q$ and/or $p$ are {\em not} identical, have one, and only one point, i.e., $M(b)$, in common.
We illustrate this in the case of two lines with common $p$ but distinct $q$'s: $q\neq q'$.
We have then that $M(b)$ of the first equals $M'(b)$ of the second iff $bq=bq'$, which implies, for $q\neq q'$, that $b=0$: i.e., the only common point is at $b=0$,
which is consistent with having a common $p$; there is no other common point for $N$ a prime number.

Of course, two lines with the same $q$ and $p$ have all their points, $N+1$ in number, in common.

From Eq. (\ref{Pj}), the trace appearing on the LHS of Eq. (\ref{orthogon. Pj}) can be written as
\begin{eqnarray}
{\rm Tr} \left[ \hat{P}_{q,p} \; \hat{P}_{q',p'} \right]
&=& \sum_b {\rm Tr} \Big[ |M_{q,p}(b), b\rangle  
\langle M_{q,p}(b), b| M_{q',p'}(b), b\rangle  
\langle M_{q',p'}(b), b| \Big] 
\nonumber \\
&+&  \sum_{b\neq b'} {\rm Tr} 
\Big[|M_{q,p}(b), b\rangle  
\langle M_{q,p}(b), b| M_{q',p'}(b'), b'\rangle  
\langle M_{q',p'}(b'), b'| \Big] \nonumber \\
&-& \sum_b {\rm Tr} 
\Big[|M_{q,p}(b), b\rangle  
\langle M_{q,p}(b), b|\Big]   \nonumber \\
&-&\sum_{b'} {\rm Tr} 
\Big[|M_{q',p'}(b'), b'\rangle  
\langle M_{q',p'}(b'), b'|\Big]   \nonumber \\
&+& {\rm Tr}\bf{I}        \nonumber \\
&\equiv& A+B-C-C'+D \; .
\end{eqnarray}
\label{}
That $D=N$ and $C=C'=N+1$ is obvious. 
For two distinct lines, thus having one point in common, $A=1$.
For two identical lines, $A=N+1$.

Now consider $B$. We have, for $b\neq b'$, 
$
\Big|\langle M_{q,p}(b), b| M_{q',p'}(b'), b'\rangle \Big|^2
=1/N
$ ,
since the bra and ket belong to two MUB.
Since the summation in $B$ contains $(N+1)N$ terms, we find $B=N+1$.

Thus
\begin{equation}
{\rm Tr} \left[ \hat{P}_{q,p} \; \hat{P}_{q',p'} \right]
=
\left\{
\begin{array}{l}
1+(N+1) -2(N+1) +N =0 , \;\; {\rm for} \;\;  (q,p)\neq(q',p') \\ \\
(N+1)+(N+1) -2(N+1) +N = N , \;\; {\rm for} \;\;  (q,p)=(q',p').
\end{array}
\right.
\end{equation}

The result of Eq. (\ref{orthogon. Pj}) then follows.

\section{Derivation of the relation Eq. (\ref{W(K) discrete 1}) between WF and Kirkwood quasi-distribution for the discrete case}
\label{derivation of W(K) discrete}

Here we proceed in analogy with the derivation given in 
App. \ref{derivation of W(K) continuous} for the continuous case, starting from Eq. (\ref{K(WW)}).

Using the product formula, Eq. (\ref{inner-product-rule}), the Kirkwood distribution  can be written as
\begin{subequations}
\begin{eqnarray}
K_{p,q}
&=& {\rm Tr}(\hat{\rho} \; \mathbb{P}_p \mathbb{P}_q) 
\label{Kirk discr 2} \\
&=& \frac{1}{N} \sum_{q',p'=0}^{N-1} 
W_{\rho}(q',p')
W_{\mathbb{P}_p \mathbb{P}_q}(q',p')  \; .
\label{K(WW) discrete}
\end{eqnarray}
\label{Kirk K(WW) discrete}
\end{subequations}
For the second WT we have 
\begin{subequations}
\begin{eqnarray}
W_{\mathbb{P}_p \mathbb{P}_q}(q',p')
&=& {\rm Tr}\left(\mathbb{P}_p \mathbb{P}_q \hat{P}_{q',p'}\right)  \\
&=& \langle p|q\rangle  \langle q|\hat{P}_{q',p'}|p\rangle  \\
&=& \sum_{\bar{q}}\langle p|q\rangle  
\langle q|\hat{P}_{q',p'}|\bar{q} \rangle \langle \bar{q}|p\rangle \; 
\label{WPpPq c}
\end{eqnarray}
and substituting the result (\ref{m. els. Pqp}) for the matrix element of the line operator, we find
\begin{eqnarray}
W_{\mathbb{P}_p \mathbb{P}_q}(q',p')
= \frac1N 
\left\{
\delta_{qq'} - \delta_{2q, \; 2q'+1}
+ e^{-\frac{2\pi i}{N}(p-p')\left[2(q-q')-1 \right]}
\right\} \; .
\label{WPpPq d}
\end{eqnarray}
\end{subequations}

From Eq. (\ref{K(WW) discrete}) we construct the combination
\begin{subequations}
\begin{eqnarray}
\sum_{\bar{q}\bar{p}} K_{\bar{p} \bar{q}} 
e^{\frac{2\pi i}{N}2(q-\bar{q})(p - \bar{p})}
&=& \frac1N \sum_{q'p'\bar{q}\bar{p}}
W_{\hat{\rho}}(q',p')
W_{\mathbb{P}_{\bar{p}} \mathbb{P}_{\bar{q}}}(q',p') 
e^{\frac{2\pi i}{N}2(q-\bar{q})(p - \bar{p})} \; .
\nonumber \\
\end{eqnarray}
Inserting the result (\ref{WPpPq d}), we find
\begin{eqnarray}
&=& \frac{1}{N^2}\sum_{q'p'} W_{\hat{\rho}}(q',p')
\sum_{\bar{q} \bar{p}}
\left\{
\delta_{\bar{q}q'} - \delta_{2\bar{q}, \; 2q'+1}
+ e^{-\frac{2\pi i}{N}(\bar{p}-p')\left[2(\bar{q}-q')-1 \right]}
\right\}
e^{\frac{2\pi i}{N}2(q-\bar{q})(p-\bar{p})} \; .
\nonumber \\
\end{eqnarray}
Evaluating the various sums we obtain Kronecker deltas, thus giving
\begin{eqnarray}
&=& \frac{1}{N^2}\sum_{q'p'} W_{\hat{\rho}}(q',p')
\left[
N \delta_{qq'} - N \delta_{2(q-q'), \; 1}
+N^2 \delta_{p p'} \delta_{2(q-q'), \; 1}
\right] \; .
\end{eqnarray}
\end{subequations}
Recalling that $1/2=(N+1)/2 \; {\rm Mod}[N]$,
\begin{eqnarray}
&&\sum_{\bar{q}\bar{p}} K_{\bar{p} \bar{q}} 
e^{\frac{2\pi i}{N}2(q-\bar{q})(p - \bar{p})}
\nonumber \\
&& \hspace{5mm}= \frac{1}{N}\sum_{p'}W_{\hat{\rho}}(q,p')
- \frac{1}{N}\sum_{p'}W_{\hat{\rho}}\big(q-(N+1)/2,p'\big)
+ W_{\hat{\rho}}\big(q-(N+1)/2, p \big) \; ,
\nonumber \\
\end{eqnarray}
or 
\begin{eqnarray}
W_{\hat{\rho}}\big(q, p \big)
&=& \langle q| \hat{\rho}|q\rangle
-\Big\langle q+(N+1)/2 \Big|\hat{\rho}\Big|q+(N+1)/2 \Big\rangle
\nonumber \\
&& \hspace{5mm} +\sum_{\bar{q}\bar{p}} K_{\bar{p} \bar{q}} 
e^{\frac{2\pi i}{N}2(q-\bar{q}+(N+1)/2)(p - \bar{p})} \; ,
\end{eqnarray}
which is the desired relation (\ref{W(K) discrete 1}).




\end{document}